\documentclass[letterpaper]{article}
\pdfoutput=1
\usepackage{jheppub}
\usepackage{amsmath}
\usepackage{graphicx}
\usepackage[config=altsf]{subfig}
\usepackage{feynmf}
\usepackage{dsfont}
\usepackage{array}
\usepackage{slashed}
\usepackage{afterpage}
\usepackage{appendix}
\usepackage{multirow}

\title{A Couplet from Flavored Dark Matter}
\author[a]{Prateek Agrawal,}
\author[b]{Zackaria Chacko,}
\author[c]{Can Kilic,}
\author[b]{and Christopher B. Verhaaren}
\affiliation[a]{Fermilab, P.O. Box 500, Batavia, IL 60510, USA}
\affiliation[b]{Maryland Center for Fundamental Physics, 
Department of Physics, \\ 
University of Maryland, College Park, MD 20742-4111}
\affiliation[c]{Theory Group, Department of Physics and Texas Cosmology Center,\\
The University of Texas at Austin, Austin, TX 78712}

\abstract { We show that a couplet, a pair of closely spaced photon
lines, in the X-ray spectrum is a distinctive feature of lepton
flavored dark matter models for which the mass spectrum is  dictated
by Minimal Flavor Violation. In such a scenario, mass splittings
between different dark matter flavors are determined by Standard Model
Yukawa couplings and can naturally be small, allowing all three
flavors to be long-lived and contribute to the observed abundance. 
Then, in the
presence of a tiny source of flavor violation, heavier dark matter
flavors can decay via a dipole transition on cosmological timescales,
giving rise to three
photon lines. The ratios of the line energies are completely
determined in terms of the charged lepton masses, and constitute a
firm prediction of this framework.  For dark matter
masses of order the weak
scale, the couplet lies in the keV--MeV region, with a much weaker
line in the eV--keV region. This scenario constitutes a potential
explanation for the recent claim of the observation of a 3.5 keV line.
The next generation of X-ray telescopes may have the necessary
resolution to resolve the double line structure of such a couplet.
}

\preprint{\\UTTG-06-15\\
TCC-001-15\\
FERMILAB-PUB-15-059-T}

\begin{document}

\maketitle

{\flushright
\underline{Haiku for a couplet}\\
\emph{
better than one line\\
is to find three, five, or nine\\
we'd settle for two}
\\
}
\section{Introduction}
%
Increasingly precise cosmological measurements indicate that about 80\% 
of the matter density of the universe is composed of constituents that 
are non-baryonic, and neutral under both color and electromagnetic 
interactions. However, the precise nature of the particles that make up 
this dark matter (DM) remains a mystery. One theoretically appealing 
possibility is that DM is composed of Weakly Interacting Massive 
Particles (WIMPs), particles with mass of order the weak scale that have 
interactions of weak scale strength with the standard model (SM) fields. 
This scenario is compelling because, provided the WIMPs were in thermal 
equilibrium with the SM at early times, just enough of them survive 
today as thermal relics to account for the observed dark matter density.

While the WIMP framework requires that DM have interactions of weak 
scale strength with the SM fields, efforts to produce it at high energy 
colliders have proven fruitless so far.  Likewise, efforts to directly 
detect DM in the laboratory through its scattering off nucleons, in 
spite of the increased sensitivity of current experiments, have also 
been unsuccessful. There are some tentative hints from indirect
detection of DM annihilation to SM today, but there is no conclusive signal.
In the wake of these null results, DM scenarios that 
retain the cosmological success of the WIMP framework while satisfying 
the current experimental bounds have become increasingly compelling.

The matter fields of the SM ($Q, U^c, D^c, L$ and $E^c$) are known to 
come in three copies, or flavors. Different flavors carry the 
same charges under the SM gauge groups, but have different couplings to 
the Higgs, and so differ in their masses. One interesting possibility, 
which has been receiving increased attention, is that DM, like the SM 
matter fields, also comes in three flavors
~\cite{Batell:2011tc,Agrawal:2011ze,
Kumar:2013hfa,Lopez-Honorez:2013wla,Kile:2013ola,Batell:2013zwa,
Agrawal:2014una, Agrawal:2014aoa,Hamze:2014wca,Lee:2014rba,Kilic:2015vka}
or has flavor-specific couplings to the 
SM~\cite{Kile:2011mn,Kamenik:2011nb,Zhang:2012da,Kile:2014jea}.
Specific DM candidates of 
this type include sneutrino DM in supersymmetric extensions of the SM, 
and Kaluza-Klein neutrino DM in theories with a universal extra 
dimension.

In~\cite{Agrawal:2011ze}, the simplest theories of flavored dark matter 
(FDM) were classified, and labelled as lepton flavored, quark flavored 
or internal flavored, based on the form of the interactions of the dark 
matter candidate with the SM fields. Models in which 
the DM has tree level interactions with the SM 
leptons but not with the quarks -- as in lepton FDM -- can naturally
account for the observed 
abundance of DM while remaining consistent with all experimental
bounds~\cite{Bai:2014osa,Chang:2014tea,Agrawal:2014ufa}. 
The reason is that strong production at a hadron collider or scattering off a 
nucleus rely on DM-quark interactions, which are loop suppressed in this 
scenario. In addition, because the average number of photons generated 
in DM annihilation to hadrons is much larger than in the case of DM 
annihilation to leptons, the limits from indirect DM searches using 
gamma rays are also weaker. Some other indirect signals of DM 
annihilation, such as the positron flux, are enhanced for lepton FDM, 
but regions of parameter space for which the DM is a thermal relic
remain viable.

In general the couplings of lepton FDM violate the flavor symmetries of 
the SM. In order to avoid conflict with the very stringent bounds on 
flavor violating processes such as $\mu \rightarrow e \gamma$, while 
giving rise to an annihilation cross section of weak scale strength, the 
couplings of DM to the SM leptons must be aligned with the SM Yukawa 
interactions. In theories where the flavor structure is consistent with 
Minimal Flavor Violation (MFV), so that the only sources of flavor 
violation are associated with the SM Yukawa couplings, this condition is 
automatically satisfied. Then each dark matter flavor is 
associated with a corresponding lepton flavor. It is this class of 
theories that we shall focus on.

In realizations of FDM that respect MFV, the mass splitting between a
pair of 
different DM flavors is dictated by the corresponding SM Yukawa couplings. In simple
models, this splitting is proportional to the 
difference in the squares of the Yukawa 
couplings, so that for a Dirac fermion we obtain,
\begin{align}
  m_{\chi, i} - m_{\chi, j} \simeq 
  \eta\,
  (y_i^2 - y_j^2) \,.
  \label{splitting1} 
\end{align}
 In this expression the index $i = 1,2,3$ runs over the three lepton 
flavors $e, \mu, \tau$. While $m_{\chi, i}$ represents the mass of the
$i^{th}$ DM flavor,
$y_i$ represents the Yukawa coupling
corresponding to the $i^{th}$ lepton flavor. 
The constant $\eta$ has the dimensions of mass and depends on the
dynamics which UV completes flavor at some high
scale $\Lambda$. 
If threshold effects at this scale generate mass splittings at tree
level, $\eta$
can naturally be of order $m_\chi$, where $m_\chi$ is the average DM mass.
In this case the largest splitting, that between the $e$ and $\tau$
flavors, is expected to be in the MeV--GeV range for 
weak scale DM. 
If tree-level contributions at the threshold are absent,
the leading effects are then loop suppressed.  The largest splitting
is then much smaller, in the keV--MeV range.

Since the Yukawa
couplings are small, the splittings between the different DM flavors are 
suppressed relative to the mass of each flavor. If the splittings are 
smaller than the electron mass, the dominant flavor-conserving decay mode
\begin{align}
  \chi_i \rightarrow \chi_j + \nu_i + \bar{\nu}_j
\end{align} 
is slow on cosmological timescales, so that the lifetimes of the
heavier flavors are much longer than the age of the universe. Then
all three flavors are expected to contribute to the observed 
DM abundance.

Now, suppose a tiny additional source of explicit 
flavor breaking is present in the theory, so that the flavor violating 
decays 
\begin{align}
  \chi_i \rightarrow \chi_j + \gamma
\end{align}
can occur on cosmological timescales and dominate over the
flavor-conserving decay. The monochromatic photons 
produced in such decays then constitute a striking signal of DM. 
Provided this new source of flavor violation is too small to contribute 
significantly to the splittings between the different DM flavors, the 
frequencies of the resulting gamma ray lines depend on the SM Yukawa 
couplings as in eq.~\eqref{splitting1}. The constant of proportionality in 
eq.~\eqref{splitting1} cancels out when ratios of frequencies are 
considered. For example, if the $\tau$ flavor of DM is the heaviest
and $\chi_e$ the lightest, we 
have,
\begin{equation} 
  \frac{\omega \left(\chi_\tau \rightarrow \chi_\mu \right)}
  {\omega \left(\chi_\tau \rightarrow \chi_e \right)} 
  =
  \frac{m_{\tau}^2 - m_{\mu}^2}{m_{\tau}^2 - m_{e}^2}
  \approx 
  1 - \frac{m_{\mu}^2}{m_{\tau}^2}
\end{equation}
Similarly,
\begin{align}
  \frac{\omega \left(\chi_\mu \rightarrow \chi_e \right)}
  {\omega \left(\chi_\tau \rightarrow \chi_e \right)}
  &=
  \frac{m_{\mu}^2 - m_{e}^2}{m_{\tau}^2 - m_{e}^2}
  \approx
  \frac{m_{\mu}^2}{m_{\tau}^2}
  \label{eq:couplet}
\end{align}
We see that this scenario predicts a pair of very closely spaced lines 
in the keV-MeV region corresponding to the $\chi_\tau \rightarrow 
\chi_e$ and $\chi_\tau \rightarrow \chi_\mu$ transitions (the 
``couplet''), as well as an isolated line in the eV-keV region. 
Remarkably, the ratios of the frequencies of these lines are a firm 
prediction of this scenario.

In the next section we review the MFV framework for models of lepton
flavored DM. In section \ref{sec:benchmark} we choose a simple
benchmark to illustrate the
phenomenology of this scenario, focusing on constraints from various
dark matter experiments and potential signals. We conclude in section
\ref{sec:conclusions}.

\section{The Framework}
\label{sec:symmetry}

In this section we study the restrictions that MFV places on the 
parameters of theories of lepton FDM. The SM has an approximate $U(3)^5$ 
flavor symmetry that acts on the three generations of fermions $Q, U^c, 
D^c, L$ and $E^c$. This symmetry is explicitly broken by the SM Yukawa 
couplings. In extensions of the SM that respect MFV, any new parameter 
that violates the SM flavor symmetries must be aligned with the SM 
Yukawa couplings. Specifically, the Yukawa couplings of the SM are 
promoted to spurions that transform under the $U(3)^5$ flavor symmetry. 
Any new interactions are then required to be invariant under this 
spurious symmetry.

In the lepton sector of the SM there is an approximate 
$U(3)_L\times U(3)_E$ flavor symmetry that acts on 
the left-handed $SU(2)$ doublets $L^A$ and $SU(2)$ singlets $E^c_i$. We 
denote $U(3)_L$ indices by capital Latin letters and 
$U(3)_E$ indices by lowercase Latin letters. This symmetry is 
violated by the SM Yukawa interactions,
 \begin{equation}
\mathcal{L}_{\text{Y}}=Y_{A}^{\phantom{A}i}L^AH^{\dag}E^c_i +\text{h.c. } 
\label{eq:smlepyuk}
 \end{equation} 
 We can, however, make the Yukawa interactions formally invariant under 
the flavor symmetry by promoting the matrix $Y_{A}^{\phantom{A}i}$ to be 
a spurion that transforms as $(\mathbf{3},\mathbf{\bar{3}})$ under 
$SU(3)_L\times SU(3)_E$ subgroup of $U(3)_L\times U(3)_E$, and has
appropriate charges under the $U(1)$ factors.

Theories of flavored DM posit a $U(3)_{\chi}$ flavor symmetry that acts 
on the DM field $\chi^{\alpha}$. We use lowercase Greek indices to 
denote the DM flavor. We focus on the case where the DM field is a 
fermion that transforms as a singlet under the SM gauge interactions, 
and has renormalizable couplings to the to the $SU(2)$ singlet leptons 
$E_i^c$ through a scalar mediator $\phi$. The mediator does not 
transform under the SM and DM flavor groups. The relevant interaction 
takes the form
 \begin{equation}
\mathcal{L}_{\lambda}=\lambda_{\alpha}^{\phantom{\alpha}i}
\chi^{\alpha}E^c_i\phi +\text{h.c. .} \label{eq:genDMint}
 \end{equation} 
 This interaction explicitly violates the $U(3)_E \times
U(3)_{\chi}$ symmetry. In general, it will give rise to lepton flavor 
violating processes at one loop. However, when MFV is imposed, the form 
of this interaction is restricted, with the result that flavor violating 
processes are forbidden. 

We impose MFV by identifying the DM flavor symmetry $U(3)_\chi$ with the 
$U(3)_{E}$ flavor symmetry that acts on the $SU(2)$ singlet leptons in 
the SM\footnote{One could also identify $U(3)_\chi$ with the
  $U(3)_{L}$ symmetry that acts on the $SU(2)$ doublet leptons (see 
  \cite{Agrawal:2011ze}), but the main results do not depend on this 
choice.}, and requiring that the new interaction be invariant under the 
spurious $U(3)^{5}$ symmetry. This leads to a restriction on the form of
$\lambda_{\alpha}^{\phantom{\alpha}j} \equiv \lambda_{i}^{\phantom{i}j}$.
To leading order in $Y_{A}^{\phantom{A}i}$ we then have,
 \begin{equation}
 \lambda_i^{\phantom{i}j} =
\widehat{\lambda}\delta_i^{\phantom{i}j}+\widetilde{\lambda} Y^{\dag A}_i Y_{A}^{\phantom{A}j}
\label{eq:diracmfvparam1}
 \end{equation}
 MFV also governs the form of the DM mass matrix. If $\chi$ is a Dirac 
fermion we can write the mass term in the Lagrangian as
 \begin{equation}
\mathcal{L}_{\text{M}}=m_{\alpha}^{\phantom{\alpha}\beta}\overline{\chi}_\beta\chi^\alpha 
\equiv m_{i}^{\phantom{i}j}\overline{\chi}_j\chi^i
\;.
\label{eq:diracDMmass}
 \end{equation}
 MFV restricts the mass matrix to be of the form  
 \begin{equation}
m_i^{\phantom{i}j} =
\widehat{m}\delta_i^{\phantom{i}j}+\widetilde{m} Y^{\dag A}_i Y_{A}^{\phantom{A}j}
\label{eq:diracmfvparam2}
 \end{equation}
 Then, in a basis where the lepton mass matrix is diagonal, we see that the 
splittings between the different DM mass eigenstates are governed by the 
SM Yukawa couplings in accordance with eq.~\eqref{splitting1}.

\begin{figure}[tp]
  \begin{center}
    \includegraphics[width=0.4\textwidth]{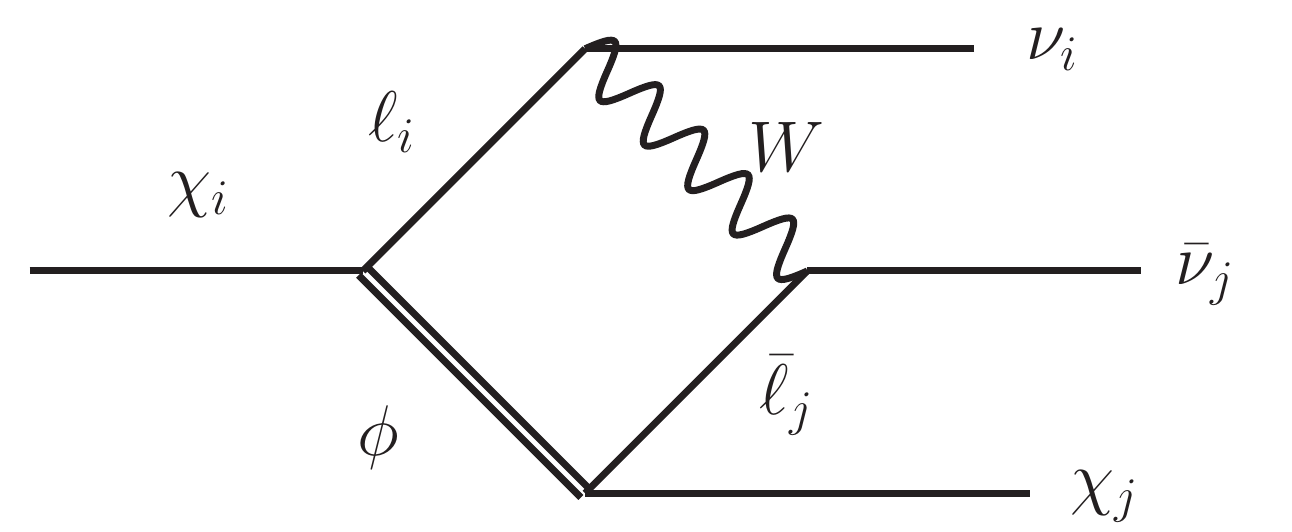}
    \includegraphics[width=0.4\textwidth]{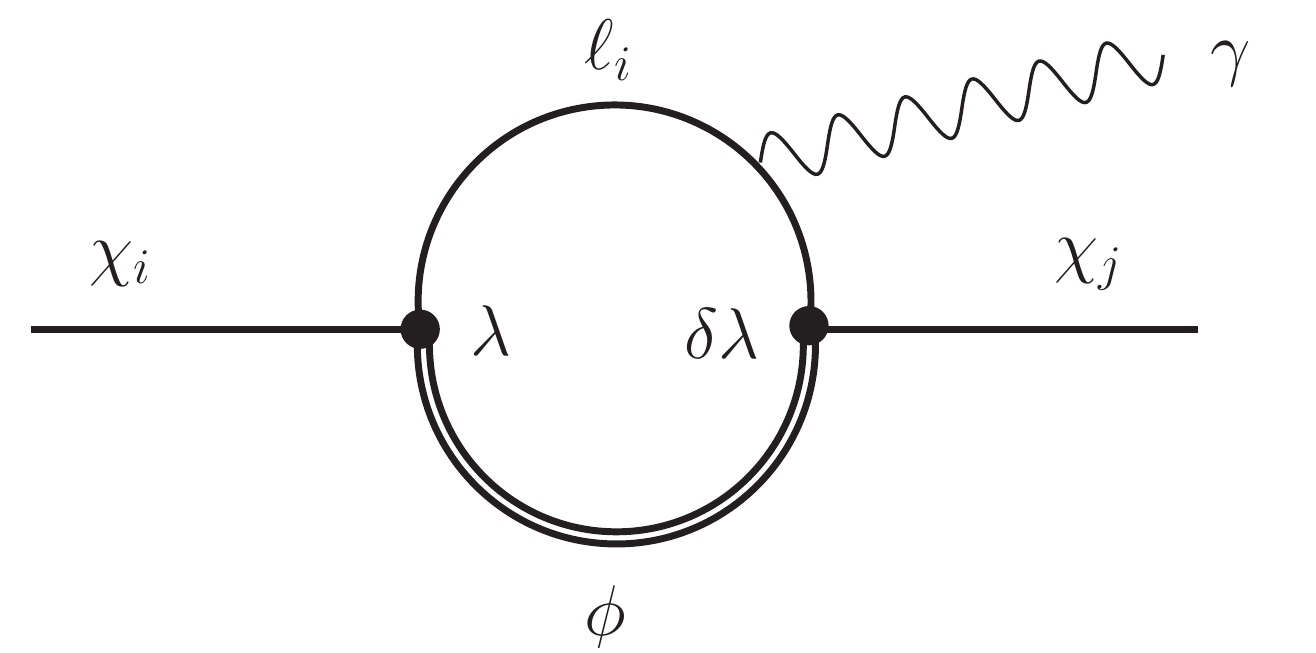}
  \end{center}
  \caption{a) Flavor preserving decay $\chi_{i}\rightarrow\chi_{j}\nu_{i}\bar{\nu}_{j}$ b) A Feynman diagram contributing to the flavor violating decay $\chi_i\rightarrow \chi_{j}\gamma$}
 \label{fig:decays}
\end{figure}
 
\section{A Benchmark Model}
\label{sec:benchmark}
In this section we construct a simplified benchmark model which exhibits
the features described above and consider its phenomenology in detail.
We choose the dark matter $\chi$ to be a Dirac
fermion which transforms as a fundamental under the SM flavor group
$SU(3)_{E}$.  The only additional terms in the Lagrangian 
beyond the SM are given by eqs.~\eqref{eq:genDMint} and
\eqref{eq:diracDMmass}.
We assume that the structure of the coupling and mass
terms is restricted by MFV, and follows
eqs.~\eqref{eq:diracmfvparam1} and \eqref{eq:diracmfvparam2}.  For
this
benchmark we will further restrict to the special case where the DM
mass terms and couplings generated at the high flavor scale $\Lambda$ are
universal, so that $\tilde{m}$ and $\tilde{\lambda}$ are zero at this
scale. The mass and interaction terms for the DM in four-component
notation at scale $\Lambda$ are then given by,
\begin{align}
\mathcal{L}
&=
m_\chi \overline{\chi}_i \chi^i
+
\left[
\frac{\lambda}{2} \,
\bar{\chi}_{i}(1+\gamma^5) e^i \phi^\dagger +\text{h.c.} 
\right]\, .
  \label{eq:lagrangian}
\end{align}
Here $e^i$ is the four-component spinor corresponding to the
charged leptons of the SM.
The only free parameters in this model are the masses of the
dark matter and the mediator, and a coupling $\lambda$. As we shall
see, the mass
splittings generated in this case are
finite at one loop and do not contain divergent pieces.
Later we will introduce a tiny source of flavor violation.

In what follows we determine the splittings between the different dark
matter flavors. We then compute expressions for the lifetimes of the heavier
flavors in the presence of a small source of flavor violation.  This
model is constrained by a number of experiments. Constraints
from $g-2$ of the muon and monophoton searches tend to be subdominant
to direct detection constraints~\cite{Agrawal:2014ufa,
Freitas:2014pua}. LHC constraints on the production
of the mediator $\phi$ and indirect
detection constraints from dark matter annihilations into
positrons and photons can also be relevant.
We study these signals in turn.

\begin{figure}[tp]
  \begin{center}
    \includegraphics[width=0.5\textwidth]{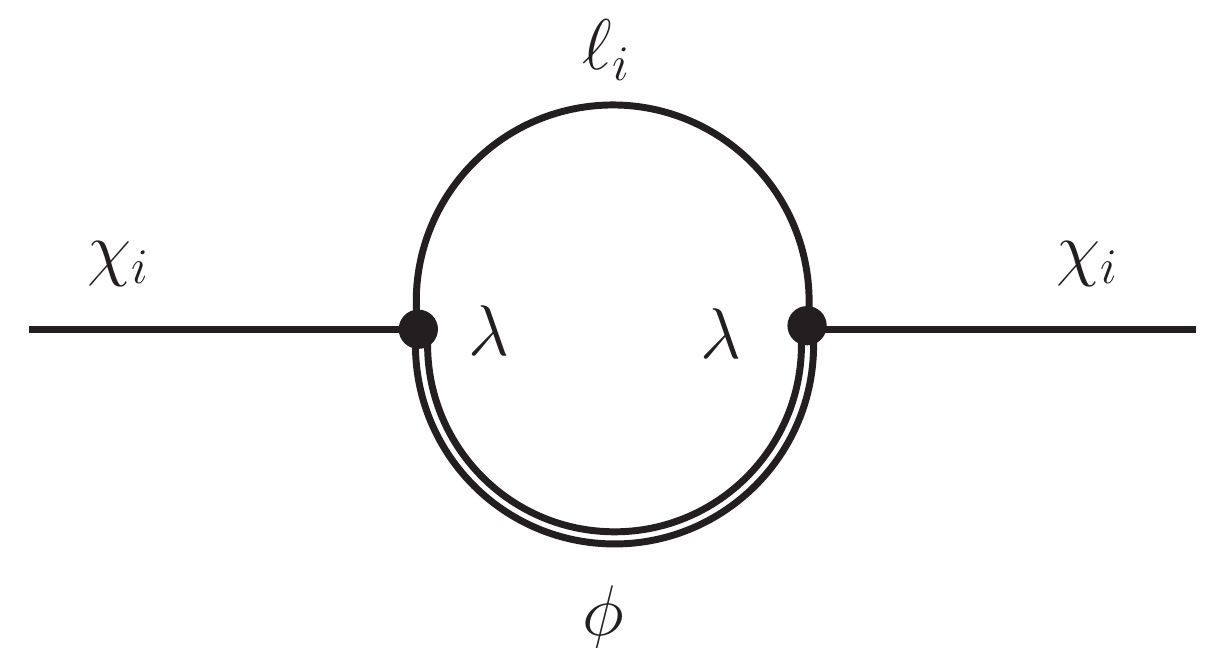}
  \end{center}
  \caption{One-loop correction to the two-point function of $\chi$}
 \label{fig:2ptfunc}
\end{figure}

\subsection{Splitting and decays}
The breaking of the lepton flavor
symmetry $U(3)\rightarrow U(1)^{3}$ by the SM lepton Yukawa couplings
is communicated to the DM sector through the 
FDM interaction, eq.~\eqref{eq:genDMint}. 
Even though they are assumed to be degenerate at tree level, 
mass splittings between the three $\chi$ flavors will be induced
at the loop level.
In particular, let us consider the 2-point function for the three flavors
of $\chi$ (see figure \ref{fig:2ptfunc}). The one-loop contributions
are identical in the limit of massless leptons, but at order
${\mathcal O}(m_{\ell}^{2})$ they begin to differ, thereby giving rise to
differences in the wavefunction renormalizations for the $\chi_{i}$.
Due to the chiral nature of the FDM coupling, it is easy to see that
there is no direct mass renormalization. Once the fields
$\chi_{i}$ are brought back to canonical normalization, however, a mass
splitting is induced between them, 
\begin{align}
  m_{\chi,i}-m_{\chi,j}
  &\equiv
  \Delta m_{ij}
  =
  \frac{m_{\chi}\lambda^{2}}{32\pi^{2}}
  \int_{0}^{1}dx\,x\,
  \log \left(
  \frac{x\,m_{\phi}^{2}+(1-x)m_{\ell,i}^{2}-x(1-x)m_{\chi}^{2}} 
  {x\,m_{\phi}^{2}+(1-x)m_{\ell,j}^{2}-x(1-x)m_{\chi}^{2}}
  \right).
\end{align}
To leading order in the Yukawa couplings this yields,
\begin{align}
\frac{\Delta m_{ij}}{m_{\chi}}
\simeq
\frac{\lambda^{2}(y_{\ell,i}^{2}-y_{\ell,j}^{2})}
{64\pi^{2}}\frac{v^{2}}{m_{\phi}^{2}}
F({m_\chi^2}/{m_\phi^2}),
\end{align}
where $y_{\ell}$ denote the Yukawa couplings of each lepton and $v =
246$ GeV is
the Higgs vacuum expectation value. The loop function $F(x)$ is given
by,
\begin{align}
  F(x)
  &=
  -\frac{x+\log(1-x)}{x^2}\simeq \frac12 + \frac{x}{3}
  +\mathcal{O}(x^2)
\end{align}
We see that $\chi_{\tau}$ is split significantly further from
$\chi_{e}$ and $\chi_{\mu}$ than these two states are
split from each other. For an overall mass scale $m_{\chi}$ in the GeV
regime and $m_\phi \sim \mathcal{O}(100)$ GeV, 
\begin{align}
  m_{\chi,\tau} - m_{\chi,\mu}
  \simeq
  m_{\chi,\tau} - m_{\chi,e}
  &\equiv \Delta m \sim \mathrm{keV}\\
  m_{\chi,\mu} - m_{\chi,e}
  &\equiv \delta m \sim \mathrm{eV} \,.
  \label{eq:split}
\end{align}
It is interesting to note that the one-loop splitting calculated above
is a finite effect, suppressed by $v^2 / m_\phi^2$ for large $\phi$
masses. Note that the sign of $\Delta m$ is not arbitrary, and as a
consequence $\chi_\tau$ is the heaviest DM flavor. At two loops there
arises a logarithmically divergent contribution to the mass splitting,
where the log divergence is cut off at the UV flavor scale, $\Lambda$.
This two-loop effect
can become  important for very large $m_\phi$ and $\Lambda$. We
estimate
that it is subleading to the finite one-loop calculation
for the range of $m_\phi$ we are interested in, provided the new physics
scale $\Lambda$ is less than 100 TeV.

Once the $\chi$ masses are split by these loop corrections, only the
lightest $\chi$ is truly stable. This is true even in the exact MFV
limit where the $U(1)^{3}$ flavor symmetry is preserved. For this
benchmark, the splittings are smaller than the mass of the electron.
Then, the leading
contribution for flavor-preserving $\chi$ decays arises at one loop
and is illustrated in the
left panel of figure \ref{fig:decays}. Note that this
contribution is very suppressed due to the following three factors.
\begin{itemize}
  \item With
a $\chi$ mass splitting of order keV, the kinematically
available phase space is extremely small. This results in a
significant suppression for the 
$1\rightarrow3$ process.
\item The loop amplitude is suppressed by the
momentum-exchange scale, or more concretely by $(\Delta m /m_{\phi})$. 
\item The lepton propagators in the loop couple to
$\phi$ on one end and to $W^{\pm}$ on the other. However, the former
couples to right-chirality leptons while the latter couples to
left-chirality leptons. Therefore both lepton propagators need a mass
insertion to obtain a nonzero amplitude, so the decay rate
is further suppressed by $m^{2}_{\ell_{i}}m^{2}_{\ell_{j}}$. 
\end{itemize}
As a consequence of these effects, heavier flavors are long-lived on
cosmological time scales.

Since the rates of flavor-preserving $\chi$ decays are so extremely
small, even a very small flavor-violating contribution can easily be
the dominant channel for the
decays of the heavier $\chi$. We add the following source of flavor
violation to the Lagrangian,
\begin{align}
  \mathcal{L}_{FV}
&=
\frac12 \delta \lambda_{ij}\,
\bar{\chi}_{i}(1+\gamma^5)e_j \phi^\dagger +\text{h.c.} 
  \label{eq:flavor-vio}
\end{align}
The leading mechanism for
flavor-violating $\chi$ decays are dipole transitions
$\chi_{i}\rightarrow\chi_{j}\gamma$, which are illustrated in the right
panel of figure \ref{fig:decays}. Note that unlike the
flavor-preserving decays, these are two-body decays, and there are no
suppressions due to lepton masses. The rate for these flavor-violating
decays can be calculated in a straightforward manner. If we assume
that all off-diagonal couplings in $\delta\lambda_{ij}$ are of the same size
$\delta\lambda\ll\lambda$, then to leading order
\begin{equation}
\Gamma_{ij}\equiv\Gamma_{\chi_{i}\rightarrow\chi_{j}\gamma}=\frac{e^{2}\lambda^{2}\delta\lambda^{2}}{1024\,\pi^{5}}\frac{(\Delta m_{ij})^{3}m_{\chi}^{2}}{m_{\phi}^4}\,,
\end{equation}
where we have neglected higher order terms proportional to lepton
masses. There are two aspects of this decay worth mentioning. First,
$\delta \lambda$ is a free parameter that allows the
heavier flavors to decay, but is otherwise small enough to leave the phenomenology
unchanged. Second, in order to not violate the parametric prediction
of the line energies, 
$\delta \lambda$ should be small enough such that its
contributions are subdominant to the previously calculated
loop-induced splittings. This condition is not difficult to
satisfy in specific cases.

Note that $\Gamma_{ij}\propto(\Delta m_{ij})^{3}$. Thus, the
rate for the transition between $\chi_{\mu}$ and $\chi_{e}$ will be
many orders of magnitude smaller than those between $\chi_{\tau}$ and
one of these two states (which have essentially the same rate as each
other). Considering that the transition between $\chi_{\mu}$ and
$\chi_{e}$ corresponds to both a smaller energy and to a much smaller
rate, we expect it will be practically unobservable. On the other
hand, a robust prediction of this setup is that if an X-ray line
signal is observed in the keV region, then at around $1\%$ energy
resolution (more precisely, an energy resolution of the order of
$y_{\mu}^{2}/y_{\tau}^{2}$) it will reveal itself as being made up of
two line signals of comparable strength.

\subsection{Relic abundance}
The dark matter annihilation rate in each channel ($\chi_i \overline{\chi}_j \to \ell_i \overline{\ell}_j$) is given by,
\begin{align}
  \langle \sigma v \rangle
  &=
  \frac{\lambda^4 m_\chi^2 }
  {32 \pi (m_\chi^2+m_\phi^2)^2}\, .
\end{align}
Since the dark matter is a Dirac fermion, there is no p-wave or chirality suppression, and thus the annihilation cross section today is the same as in the early universe to a good approximation.

\begin{figure}[tp]
  \begin{center}
    \includegraphics[width=0.6\textwidth]{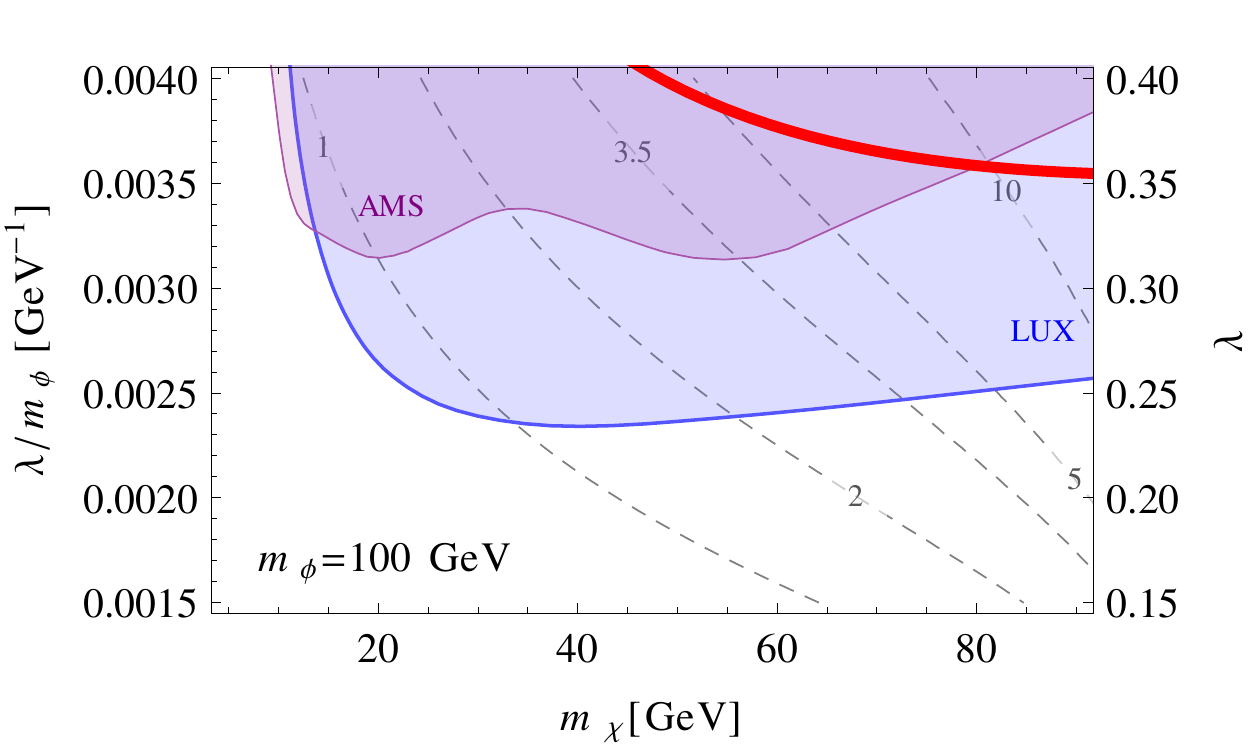}
    \\
    \includegraphics[width=0.6\textwidth]{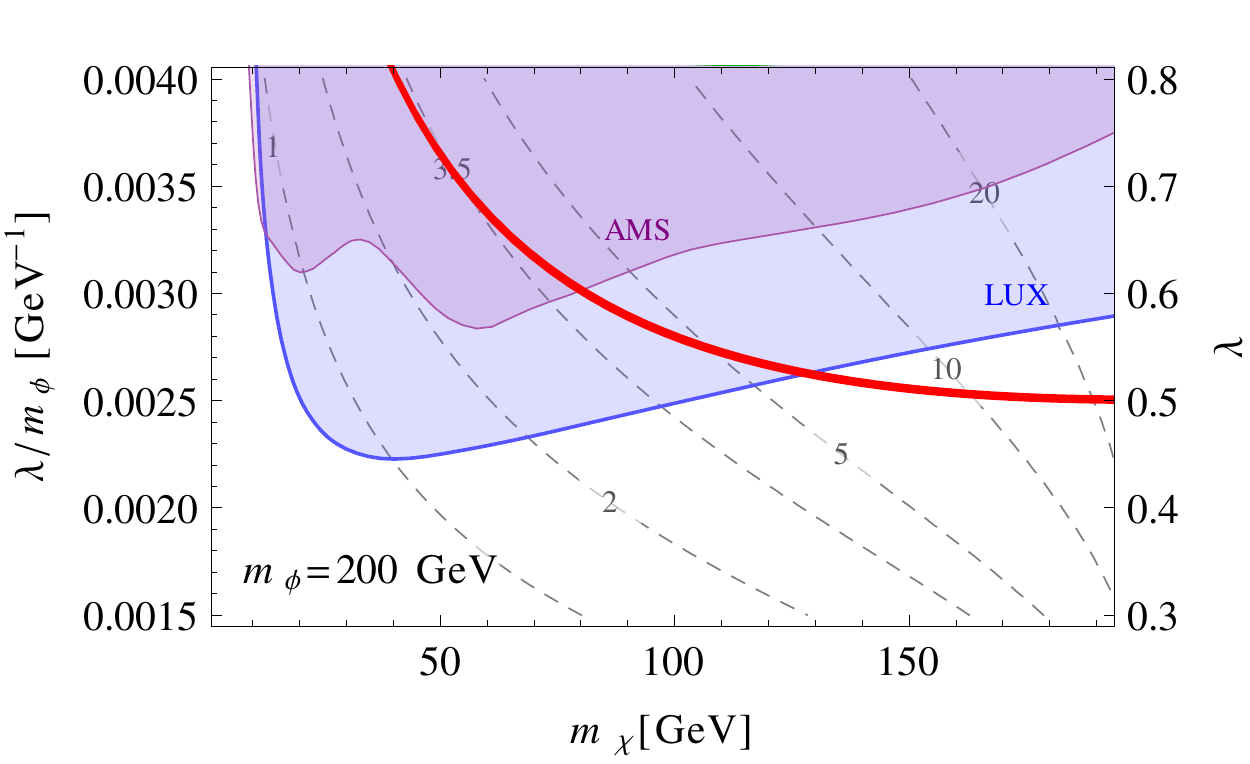}
    \\
    \includegraphics[width=0.6\textwidth]{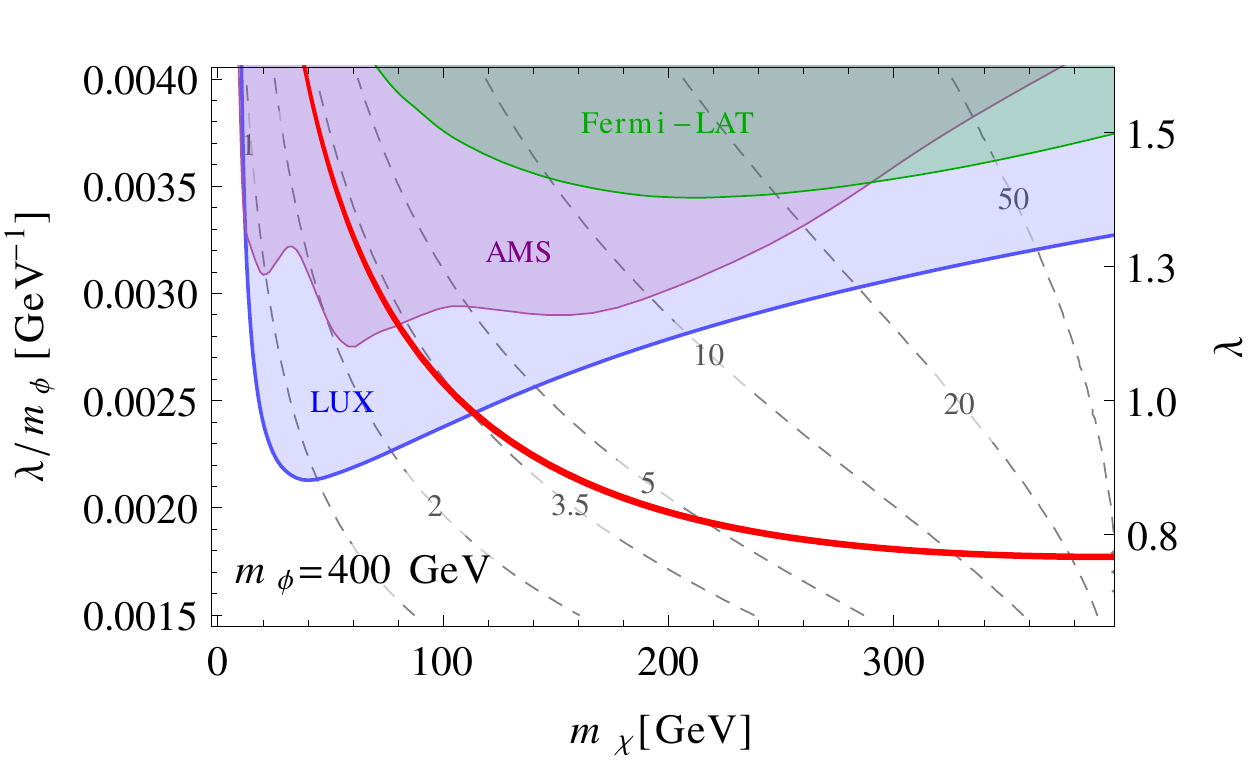}
  \end{center}
  \caption{For mediator masses $m_\phi=100,200$ and 400 GeV, we plot
  the position of the X-ray signal (in keV, gray dashed contours) as
  well as a number of constraints. Direct detection constraints from
  LUX are shown as the blue-shaded region, while the indirect
  detection constraints from positrons and photons are shown as the
  purple and green-shaded regions, respectively. 
  The red band shows the region where
  correct relic abundance is obtained. 
}
 \label{fig:constraints}
\end{figure}

Since all flavor combinations of dark matter co-annihilate with one
another with the same cross section, the cross section that gives rise
to the correct relic abundance today is the same as for a single
species of Dirac fermion DM, given by
\begin{align}
  \langle \sigma v \rangle
  &=
  2\times(2.2\times10^{-26}~{\rm cm^{3}/s}).\label{eq:factortwo}
\end{align}
The factor of two relative to the canonical quoted value (for Majorana
DM) arises due to the Dirac nature of the dark matter.
The region of parameter space leading to the correct relic density is
shown in figure~\ref{fig:constraints} as a red band.


Note that the above calculation applies only if we assume that the
interaction with leptons alone is responsible for the dark matter
thermal relic abundance. Any coupling to additional non-SM
states will alter these numbers. This constraint can also be relaxed
if the dark matter relic density is set by an asymmetry, which can
arise rather naturally in these models~\cite{Hamze:2014wca}.

\subsection{Direct detection}
As discussed in detail in \cite{Agrawal:2011ze}, lepton flavored dark
matter can scatter off nuclei via a one-loop photon exchange. These
constraints can be severe in the region where the dark matter is a
thermal relic. The dominant contribution to the WIMP-nucleon cross
section is flavor diagonal and for each flavor of FDM it is given by,
\begin{align}
  \sigma_n
  &=
  \frac{\mu_n^2 Z^2}{ A^2 \pi}
  \sum_{\ell}
  \left(
  \frac{\lambda^2 e^2}{64\pi^2 m_\phi^2}
  \left[1+\frac23 \log\frac{\Lambda_\ell^2}{m_\phi^2}
  \right]
  \right)^2
  \, .
\end{align}
Here $\mu_n$ is the reduced mass of the dark matter--nucleon system
and $\Lambda_{\ell}$ represents the infrared cutoff in the loop
calculation for the effective DM-photon coupling. This cutoff is
$m_{\ell}$, the mass of the corresponding lepton, unless $m_{\ell}$ is
smaller than the momentum exchange in the process, of the order of
10~MeV. We therefore use $\Lambda_{\tau}=m_{\tau}$ and
$\Lambda_{\mu}=m_{\mu}$, but set $\Lambda_{e}=10~$MeV. In extracting constraints
from the null results of direct detection experiments such as LUX, we
use the total rate, summed over all three FDM flavors. The region of
parameter space excluded by LUX is shown in
figure~\ref{fig:constraints} as the blue-shaded region.

\subsection{Indirect detection} 
In the limit
$m_{\chi}\ll m_{\phi}$, $(\Delta m)^{2}$ and $\langle\sigma v\rangle$
both scale approximately as $m_{\chi}^{2}\lambda^{4}/m_{\phi}^{4}$,
and therefore choosing a fixed mass splitting or requiring thermal
relic abundance leads to potentially observable signals for indirect
detection searches in photons and positrons (with the caveats
mentioned at the end of the previous paragraph). The constraints from
both indirect detection channels are more stringent for lower mass
dark matter, since the signal rate scales as the square of the $\chi$
number density, which itself scales as $m_{\chi}^{-1}$, while the
background flux as a function of energy does not change as rapidly.
Therefore, for a given $\Delta m$ or $\langle\sigma v\rangle$, these
constraints can be weakened by increasing the dark matter mass and
either decreasing the coupling or increasing $m_{\phi}$.

For the positron constraint from the AMS-02
experiment~\cite{Ibarra:2013zia}, the signal has contributions both
from the prompt positrons produced when one of the annihilating
particles is $\overline{\chi}_{e}$, and also from secondary positrons
from the decays of  $\mu^+$ and $\tau^+$ that are produced when one of
the annihilating particles is $\overline{\chi}_{\mu}$ or
$\overline{\chi}_{\tau}$. The spectrum of the secondary positrons is
shifted towards lower energies compared to the prompt positrons (not
to mention that the branching ratio of $\tau\to e+X$ is rather low),
and therefore the bound from AMS-02 comes mostly from the prompt
positrons. The bound is shown in figure~\ref{fig:constraints} as the
purple-shaded region. Note that the positron constraint is
significantly weaker than the constraint from direct detection across
the parameter space, and therefore the inclusion or non-inclusion of secondary
positrons in determining the bound turns out to be academic. Owing to
the relative factor of
two between Dirac and Majorana DM (eq. \eqref{eq:factortwo}), the
latter being relevant for SUSY for which the AMS bounds are
calculated, the bound on the FDM annihilation cross section leading
from prompt positrons (any one of the three $\chi_{\ell}$ flavors
annihilating with $\overline{\chi}_{e}$) is related to the total
annihilation cross section (all nine annihilation channels) as
\begin{align}
  \langle \sigma v \rangle 
  \leq
  6 \langle \sigma v \rangle_{bound,e^+}
  \, ,
\end{align}
where $\langle \sigma v \rangle_{bound,e^+}$ is the experimental
bound quoted in~\cite{Ibarra:2013zia} for a Majorana DM annihilating
to $e^+ e^-$ with
$100\%$ branching fraction.

Similar to the case of positron constraints being most sensitive to
prompt positrons in the final state, indirect detection in photons is
most sensitive to $\tau$'s in the final state, since more 
photons are produced from $\tau$'s than from $e$'s or $\mu$'s. One can
therefore formulate the bound from indirect detection in the photon
final state~\cite{Tavakoli:2013zva} in terms of the effective
annihilation cross section leading to the production of $\tau$'s,
\begin{align}
  \langle \sigma v \rangle
  \leq
  6 \langle \sigma v \rangle_{bound,\gamma}
  \, ,
\end{align}
where $\langle \sigma v \rangle_{bound,\gamma}$ is the experimental
bound quote in~\cite{Tavakoli:2013zva} for a Majorana DM annihilating
to $\tau^+ \tau^-$ with $100\%$
branching ratio.
The constraint from indirect detection in photons is shown in
figure~\ref{fig:constraints} as the green-shaded region. 

Dark matter annihilating to leptons can potentially have significant 
constraints from the CMB~\cite{Padmanabhan:2005es,2009PhRvD..80d3526S}. However, the annihilation into muons and
taus has a low efficiency to inject energy into the
CMB~\cite{Madhavacheril:2013cna}, and the constraints are
subdominant to the other constraints considered above.

\begin{figure}[tp]
  \begin{center}
    \includegraphics[width=0.6\textwidth]{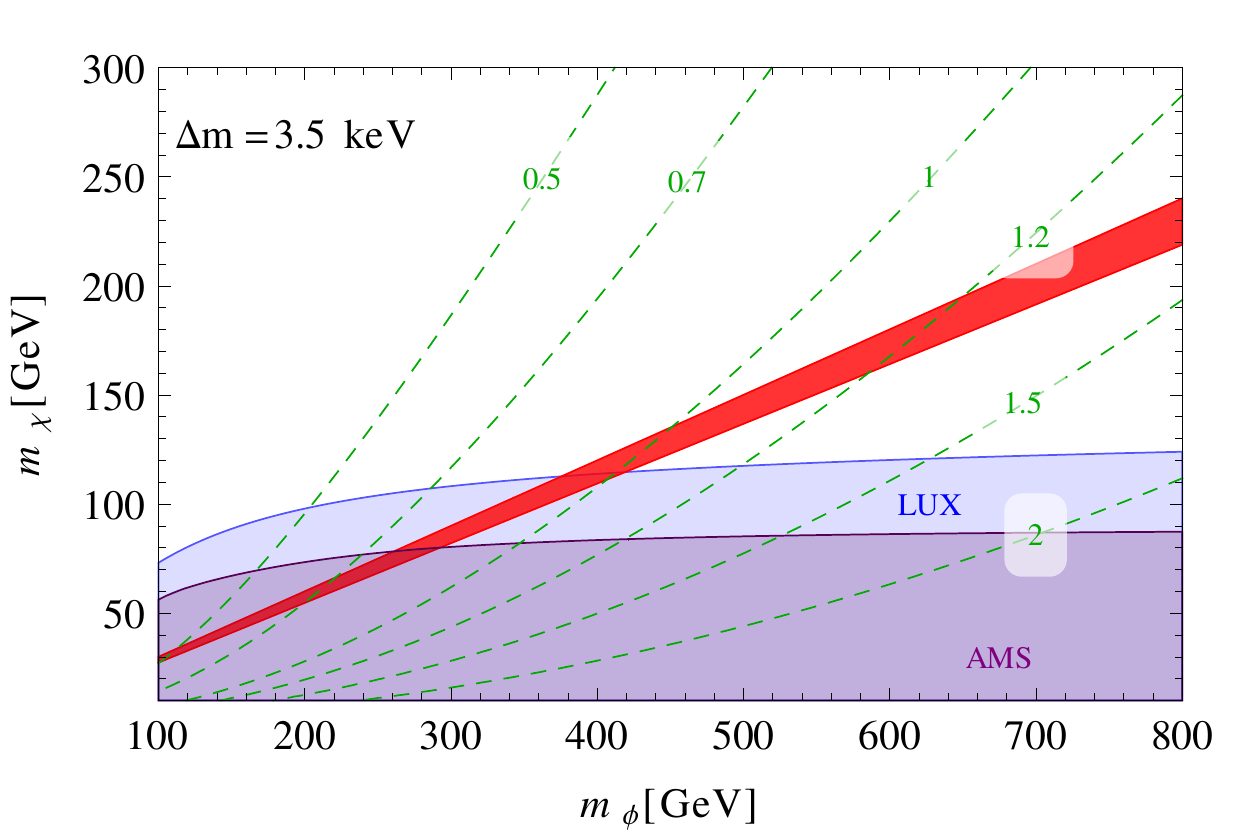}
  \end{center}
  \caption{Constraints on the mass of dark matter, $m_\chi$ and the
  mediator, $m_\phi$ when the X-ray line is at 3.5 keV. The contours
show the value of coupling $\lambda$, and the red band shows the
region where correct relic abundance is obtained. The blue and purple-shaded
regions show the exclusion from LUX and from AMS respectively.}
 \label{fig:chiphi}
\end{figure}

\subsection{The 3.5 keV line}

An X-ray line signal has been observed at
3.5~keV~\cite{Bulbul:2014sua,Boyarsky:2014jta}. 
We note that there is
currently no consensus about the interpretation of this observation
as arising from a dark matter signal~\cite{Riemer-Sorensen:2014yda,
Boyarsky:2014ska,
Jeltema:2014qfa,Malyshev:2014xqa,Anderson:2014tza,Boyarsky:2014paa,Bulbul:2014ala,Jeltema:2014mla,Iakubovskyi:2014yxa,Carlson:2014lla,
Tamura:2014mta, Urban:2014yda}. Nevertheless, a
large number of DM models have been proposed to explain this signal.
The ideas that have been proposed include 
sterile neutrinos \cite{Patra:2014sua,
Lattanzi:2014mia,
Haba:2014taa,
Shuve:2014doa,
Rodejohann:2014eka,
Abada:2014nwa,
Baek:2014qwa,
Chakraborty:2014sda,
Kang:2014cia,
Dobrynina:2014zza,
Okada:2014zea,
Okada:2014oda,
Abada:2014zra,
Ishida:2014fra,
Cline:2014eaa,
Modak:2014vva,
Robinson:2014bma,
Chakraborty:2014tma,
Adulpravitchai:2014xna,
Patra:2014pga,
Merle:2014xpa,
Frigerio:2014ifa,
Tsuyuki:2014aia,
Abazajian:2014gza,
Kang:2014mea,
Harada:2014lma}, axions \cite{Choi:2014tva,
Liew:2014gia,
Higaki:2014qua,
Kawasaki:2014una,
Higaki:2014zua,
Jaeckel:2014qea,
Lee:2014xua,
Kong:2014gea,
Lello:2014yha,
Ishida:2014dlp,
Rosner:2014cha,
Alvarez:2014gua,
Henning:2014dha,
Cicoli:2014bfa,
Dias:2014osa,
Conlon:2014xsa}, supersymmetry \cite{Hamaguchi:2014sea,
Kolda:2014ppa,
Bomark:2014yja,
Ko:2014xda,
Demidov:2014hka,
Dutta:2014saa} and a number of other mechanisms \cite{Pavlidou:2013zha,
Cline:2014vsa,
Baek:2014sda,
Falkowski:2014sma,
Schutz:2014nka,
Boddy:2014qxa,
Farzan:2014foo,
Cline:2014kaa,
Chiang:2014xra,
Geng:2014zqa,
Conlon:2014wna,
Chen:2014vna,
Nakayama:2014rra,
Baek:2014poa,
Lee:2014koa,
Babu:2014pxa,
Dudas:2014ixa,
Queiroz:2014yna,
Nakayama:2014cza,
Allahverdi:2014dqa,
Bezrukov:2014nza,
Nakayama:2014ova,
Frandsen:2014lfa,
Aisati:2014nda,
Krall:2014dba,
Finkbeiner:2014sja,
Babu:2014uoa,
Arcadi:2014dca,
Cheung:2014tha,
Baer:2014eja,
Biswas:2015sva,
Berlin:2015sia}.
We now show that the 3.5 keV line, if confirmed, can be easily
accommodated within the framework of the minimal model.
We show in figure \ref{fig:constraints} 
the region of parameter space that provides the right
relic abundance for dark matter, consistent with a 3.5 keV
line. Fixing the splitting to be $3.5$ keV, we show the direct
detection and the AMS positron constraint in figure \ref{fig:chiphi}
along with the region of parameter space consistent with the
requirement of correct relic abundance.

This scenario further predicts that closer inspection of such a line
will reveal two closely spaced lines corresponding to the $\chi_\tau
\to \chi_\mu$ 
and $\chi_\tau \to \chi_e$ transitions. As noted in eq.~\eqref{eq:couplet}, the
ratio of the line energies in
this couplet are set by the charged lepton masses. Therefore,
\begin{align}
  \delta \omega
  &=
  \omega(\chi_\tau\to\chi_e)-\omega(\chi_\tau \to \chi_\mu)
  \\
  &\approx
  \omega_0 \frac{m_\mu^2}{m_\tau^2}
  = 12.4\ \mathrm{ eV}
\end{align}
where $\omega_0 = 3.5$ keV is the average frequency of the two $\chi_\tau$
decay lines. 
The broadening of the line due to the kinetic energy of the DM
scales with its velocity.
For typical astrophysical sources, the DM velocity ranges from
{(1--3)}$\times10^{-3}$, resulting in a broadening of 
$\mathcal{O}(1$--$10)$ eV. Whether the double line
feature gets washed out by the thermal broadening thus depends on the
astrophysical source.
While this splitting is not currently measurable, it is
within the design resolution of upcoming experiments like ASTRO-H
\cite{Takahashi:2010fa,Takahashi:2014cea}.
The couplet constitutes a
``smoking gun" signal of lepton FDM scenarios at these experiments.

The lifetime for a decaying dark matter candidate to be consistent
with the observed signal is given by (see for instance~\cite{Farzan:2014foo}),
\begin{align}
  \tau_{DM}
  &\simeq
  (10^{28}\ \mathrm{sec})
  \frac{ 7\ \mathrm{keV}}{m_\chi}
  \, .
\end{align}
For $m_{\chi}=150$~GeV, one obtains $\tau_{DM}\approx10^{20}$~s.
For $m_{\phi}=500$~GeV and
$\lambda\simeq 1$, this would require
$\delta\lambda \simeq 10^{-8}$. The additional mass splittings introduced by
this level of flavor violation are subdominant to the MFV
contributions calculated above.

\section{Conclusions} 
\label{sec:conclusions}

We have studied models of lepton flavored dark matter within the 
MFV framework. In this scenario, mass
splittings between different dark matter flavors can
naturally be small enough that tree level decays of heavy flavors are
kinematically forbidden.
Then all three 
flavors of dark matter can be long-lived on cosmological time scales
(the lightest flavor being exactly stable). 
The ratios of the mass splittings
between the three possible pairings among the DM flavors are
predicted.

When even a very small source of flavor violation is present in the dark
sector, a new decay channel becomes available for the decay of
heavier dark matter flavors through a dipole transition. While the lifetime
associated with this decay may be orders of magnitude longer
than the age of the universe, it can still be the dominant decay channel and
gives rise to a very distinct final state. These decays
result in two very closely
separated photon lines, the couplet. For weak scale DM, the overall
energy of the couplet is naturally in the keV--MeV range, with the
splitting of the two lines in the eV--keV range.

We have focused on the detailed phenomenology of a specific model of lepton
flavored DM. In this model, the mass splittings are radiatively generated by
finite one loop effects arising from the breaking of the lepton flavor
symmetry by Yukawa couplings. 
The sign of the contribution is fixed, with the result that
$\chi_{\tau}$ is the heaviest and $\chi_e$ the lightest state.

This scenario is a potential explanation for the claimed observation
of a 3.5 keV line in the X-ray spectrum, and there exist regions in
parameter space of the model where such an explanation is entirely
consistent with the observed DM relic density as well as with
experimental constraints set by a number of direct and indirect
detection experiments. While a dark matter explanation of this line is
in dispute, the smoking-gun double-line
structure predicted by the couplet can be directly tested
experimentally. In particular, the next generation experiments
might be able to resolve any such feature in the X-ray
spectrum.

\acknowledgments We thank Ilias Cholis and Dan Hooper for useful
discussions.  CK would also like to thank the Aspen Center for Physics
(supported by the National Science Foundation under Grant No.
PHYS-1066293) as well as the Perimeter Institute for Theoretical
Physics (supported by the Government of Canada through Industry Canada
and by the Province of Ontario through the Ministry of Research and
Innovation), where part of this work was completed, for their
hospitality.  ZC and CV are supported by NSF under grant PHY-1315155.
CK is supported by NSF grant numbers PHY-1315983 and PHY-1316033.
Fermilab is operated by Fermi Research Alliance, LLC under Contract
No. De-AC02-07CH11359 with the United States Department of Energy.

\bibliographystyle{JHEP}
\bibliography{couplet.bib}

\providecommand{\href}[2]{#2}\begingroup\raggedright\begin{thebibliography}{10%
0}

\bibitem{Batell:2011tc}
B.~Batell, J.~Pradler, and M.~Spannowsky, {\it {Dark Matter from Minimal Flavor
  Violation}},  {\em JHEP} {\bf 1108} (2011) 038,
  [\href{http://arxiv.org/abs/1105.1781}{{\tt arXiv:1105.1781}}].

\bibitem{Agrawal:2011ze}
P.~Agrawal, S.~Blanchet, Z.~Chacko, and C.~Kilic, {\it {Flavored Dark Matter,
  and Its Implications for Direct Detection and Colliders}},  {\em Phys.Rev.}
  {\bf D86} (2012) 055002, [\href{http://arxiv.org/abs/1109.3516}{{\tt
  arXiv:1109.3516}}].

\bibitem{Kumar:2013hfa}
A.~Kumar and S.~Tulin, {\it {Top-flavored dark matter and the forward-backward
  asymmetry}},  {\em Phys.Rev.} {\bf D87} (2013), no.~9 095006,
  [\href{http://arxiv.org/abs/1303.0332}{{\tt arXiv:1303.0332}}].

\bibitem{Lopez-Honorez:2013wla}
L.~Lopez-Honorez and L.~Merlo, {\it {Dark matter within the minimal flavour
  violation ansatz}},  {\em Phys.Lett.} {\bf B722} (2013) 135--143,
  [\href{http://arxiv.org/abs/1303.1087}{{\tt arXiv:1303.1087}}].

\bibitem{Kile:2013ola}
J.~Kile, {\it {Flavored Dark Matter: A Review}},  {\em Mod.Phys.Lett.} {\bf
  A28} (2013) 1330031, [\href{http://arxiv.org/abs/1308.0584}{{\tt
  arXiv:1308.0584}}].

\bibitem{Batell:2013zwa}
B.~Batell, T.~Lin, and L.-T. Wang, {\it {Flavored Dark Matter and R-Parity
  Violation}},  {\em JHEP} {\bf 1401} (2014) 075,
  [\href{http://arxiv.org/abs/1309.4462}{{\tt arXiv:1309.4462}}].

\bibitem{Agrawal:2014una}
P.~Agrawal, B.~Batell, D.~Hooper, and T.~Lin, {\it {Flavored Dark Matter and
  the Galactic Center Gamma-Ray Excess}},  {\em Phys.Rev.} {\bf D90} (2014),
  no.~6 063512, [\href{http://arxiv.org/abs/1404.1373}{{\tt arXiv:1404.1373}}].

\bibitem{Agrawal:2014aoa}
P.~Agrawal, M.~Blanke, and K.~Gemmler, {\it {Flavored dark matter beyond
  Minimal Flavor Violation}},  {\em JHEP} {\bf 1410} (2014) 72,
  [\href{http://arxiv.org/abs/1405.6709}{{\tt arXiv:1405.6709}}].

\bibitem{Hamze:2014wca}
A.~Hamze, C.~Kilic, J.~Koeller, C.~Trendafilova, and J.-H. Yu, {\it
  {Lepton-Flavored Asymmetric Dark Matter and Interference in Direct
  Detection}},  \href{http://arxiv.org/abs/1410.3030}{{\tt arXiv:1410.3030}}.

\bibitem{Lee:2014rba}
C.-J. Lee and J.~Tandean, {\it {Lepton-Flavored Scalar Dark Matter with Minimal
  Flavor Violation}},  \href{http://arxiv.org/abs/1410.6803}{{\tt
  arXiv:1410.6803}}.

\bibitem{Kilic:2015vka}
C.~Kilic, M.~D. Klimek, and J.-H. Yu, {\it {Signatures of Top Flavored Dark
  Matter}},  \href{http://arxiv.org/abs/1501.02202}{{\tt arXiv:1501.02202}}.

\bibitem{Kile:2011mn}
J.~Kile and A.~Soni, {\it {Flavored Dark Matter in Direct Detection Experiments
  and at LHC}},  {\em Phys.Rev.} {\bf D84} (2011) 035016,
  [\href{http://arxiv.org/abs/1104.5239}{{\tt arXiv:1104.5239}}].

\bibitem{Kamenik:2011nb}
J.~F. Kamenik and J.~Zupan, {\it {Discovering Dark Matter Through Flavor
  Violation at the LHC}},  {\em Phys.Rev.} {\bf D84} (2011) 111502,
  [\href{http://arxiv.org/abs/1107.0623}{{\tt arXiv:1107.0623}}].

\bibitem{Zhang:2012da}
Y.~Zhang, {\it {Top Quark Mediated Dark Matter}},  {\em Phys.Lett.} {\bf B720}
  (2013) 137--141, [\href{http://arxiv.org/abs/1212.2730}{{\tt
  arXiv:1212.2730}}].

\bibitem{Kile:2014jea}
J.~Kile, A.~Kobach, and A.~Soni, {\it {Lepton-Flavored Dark Matter}},
  \href{http://arxiv.org/abs/1411.1407}{{\tt arXiv:1411.1407}}.

\bibitem{Bai:2014osa}
Y.~Bai and J.~Berger, {\it {Lepton Portal Dark Matter}},  {\em JHEP} {\bf 1408}
  (2014) 153, [\href{http://arxiv.org/abs/1402.6696}{{\tt arXiv:1402.6696}}].

\bibitem{Chang:2014tea}
S.~Chang, R.~Edezhath, J.~Hutchinson, and M.~Luty, {\it {Leptophilic Effective
  WIMPs}},  {\em Phys.Rev.} {\bf D90} (2014), no.~1 015011,
  [\href{http://arxiv.org/abs/1402.7358}{{\tt arXiv:1402.7358}}].

\bibitem{Agrawal:2014ufa}
P.~Agrawal, Z.~Chacko, and C.~B. Verhaaren, {\it {Leptophilic Dark Matter and
  the Anomalous Magnetic Moment of the Muon}},  {\em JHEP} {\bf 1408} (2014)
  147, [\href{http://arxiv.org/abs/1402.7369}{{\tt arXiv:1402.7369}}].

\bibitem{Freitas:2014pua}
A.~Freitas, J.~Lykken, S.~Kell, and S.~Westhoff, {\it {Testing the Muon g-2
  Anomaly at the LHC}},  {\em JHEP} {\bf 1405} (2014) 145,
  [\href{http://arxiv.org/abs/1402.7065}{{\tt arXiv:1402.7065}}].

\bibitem{Ibarra:2013zia}
A.~Ibarra, A.~S. Lamperstorfer, and J.~Silk, {\it {Dark matter annihilations
  and decays after the AMS-02 positron measurements}},  {\em Phys.Rev.} {\bf
  D89} (2014) 063539, [\href{http://arxiv.org/abs/1309.2570}{{\tt
  arXiv:1309.2570}}].

\bibitem{Tavakoli:2013zva}
M.~Tavakoli, I.~Cholis, C.~Evoli, and P.~Ullio, {\it {Constraints on dark
  matter annihilations from diffuse gamma-ray emission in the Galaxy}},  {\em
  JCAP} {\bf 1401} (2014) 017, [\href{http://arxiv.org/abs/1308.4135}{{\tt
  arXiv:1308.4135}}].

\bibitem{Padmanabhan:2005es}
N.~Padmanabhan and D.~P. Finkbeiner, {\it {Detecting dark matter annihilation
  with CMB polarization: Signatures and experimental prospects}},  {\em
  Phys.Rev.} {\bf D72} (2005) 023508,
  [\href{http://arxiv.org/abs/astro-ph/0503486}{{\tt astro-ph/0503486}}].

\bibitem{2009PhRvD..80d3526S}
T.~R. {Slatyer}, N.~{Padmanabhan}, and D.~P. {Finkbeiner}, {\it {CMB
  constraints on WIMP annihilation: Energy absorption during the recombination
  epoch}},  {\em Phys.Rev} {\bf 80} (Aug., 2009) 043526,
  [\href{http://arxiv.org/abs/0906.1197}{{\tt arXiv:0906.1197}}].

\bibitem{Madhavacheril:2013cna}
M.~S. Madhavacheril, N.~Sehgal, and T.~R. Slatyer, {\it {Current Dark Matter
  Annihilation Constraints from CMB and Low-Redshift Data}},  {\em Phys.Rev.}
  {\bf D89} (2014), no.~10 103508, [\href{http://arxiv.org/abs/1310.3815}{{\tt
  arXiv:1310.3815}}].

\bibitem{Bulbul:2014sua}
E.~Bulbul, M.~Markevitch, A.~Foster, R.~K. Smith, M.~Loewenstein, et~al., {\it
  {Detection of An Unidentified Emission Line in the Stacked X-ray spectrum of
  Galaxy Clusters}},  {\em Astrophys.J.} {\bf 789} (2014) 13,
  [\href{http://arxiv.org/abs/1402.2301}{{\tt arXiv:1402.2301}}].

\bibitem{Boyarsky:2014jta}
A.~Boyarsky, O.~Ruchayskiy, D.~Iakubovskyi, and J.~Franse, {\it {Unidentified
  Line in X-Ray Spectra of the Andromeda Galaxy and Perseus Galaxy Cluster}},
  {\em Phys.Rev.Lett.} {\bf 113} (2014), no.~25 251301,
  [\href{http://arxiv.org/abs/1402.4119}{{\tt arXiv:1402.4119}}].

\bibitem{Riemer-Sorensen:2014yda}
S.~Riemer-Sorensen, {\it {Questioning a 3.5 keV dark matter emission line}},
  \href{http://arxiv.org/abs/1405.7943}{{\tt arXiv:1405.7943}}.

\bibitem{Boyarsky:2014ska}
A.~Boyarsky, J.~Franse, D.~Iakubovskyi, and O.~Ruchayskiy, {\it {Checking the
  dark matter origin of 3.53~keV line with the Milky Way center}},
  \href{http://arxiv.org/abs/1408.2503}{{\tt arXiv:1408.2503}}.

\bibitem{Jeltema:2014qfa}
T.~E. Jeltema and S.~Profumo, {\it {Dark matter searches going bananas: the
  contribution of Potassium (and Chlorine) to the 3.5 keV line}},
  \href{http://arxiv.org/abs/1408.1699}{{\tt arXiv:1408.1699}}.

\bibitem{Malyshev:2014xqa}
D.~Malyshev, A.~Neronov, and D.~Eckert, {\it {Constraints on 3.55 keV line
  emission from stacked observations of dwarf spheroidal galaxies}},  {\em
  Phys.Rev.} {\bf D90} (2014), no.~10 103506,
  [\href{http://arxiv.org/abs/1408.3531}{{\tt arXiv:1408.3531}}].

\bibitem{Anderson:2014tza}
M.~E. Anderson, E.~Churazov, and J.~N. Bregman, {\it {Non-Detection of X-Ray
  Emission From Sterile Neutrinos in Stacked Galaxy Spectra}},
  \href{http://arxiv.org/abs/1408.4115}{{\tt arXiv:1408.4115}}.

\bibitem{Boyarsky:2014paa}
A.~Boyarsky, J.~Franse, D.~Iakubovskyi, and O.~Ruchayskiy, {\it {Comment on the
  paper "Dark matter searches going bananas: the contribution of Potassium (and
  Chlorine) to the 3.5 keV line" by T. Jeltema and S. Profumo}},
  \href{http://arxiv.org/abs/1408.4388}{{\tt arXiv:1408.4388}}.

\bibitem{Bulbul:2014ala}
E.~Bulbul, M.~Markevitch, A.~R. Foster, R.~K. Smith, M.~Loewenstein, et~al.,
  {\it {Comment on "Dark matter searches going bananas: the contribution of
  Potassium (and Chlorine) to the 3.5 keV line"}},
  \href{http://arxiv.org/abs/1409.4143}{{\tt arXiv:1409.4143}}.

\bibitem{Jeltema:2014mla}
T.~Jeltema and S.~Profumo, {\it {Reply to Two Comments on "Dark matter searches
  going bananas the contribution of Potassium (and Chlorine) to the 3.5 keV
  line"}},  \href{http://arxiv.org/abs/1411.1759}{{\tt arXiv:1411.1759}}.

\bibitem{Iakubovskyi:2014yxa}
D.~Iakubovskyi, {\it {New emission line at ~3.5 keV - observational status,
  connection with radiatively decaying dark matter and directions for future
  studies}},  \href{http://arxiv.org/abs/1410.2852}{{\tt arXiv:1410.2852}}.

\bibitem{Carlson:2014lla}
E.~Carlson, T.~Jeltema, and S.~Profumo, {\it {Where do the 3.5 keV photons come
  from? A morphological study of the Galactic Center and of Perseus}},
  \href{http://arxiv.org/abs/1411.1758}{{\tt arXiv:1411.1758}}.

\bibitem{Tamura:2014mta}
T.~Tamura, R.~Iizuka, Y.~Maeda, K.~Mitsuda, and N.~Y. Yamasaki, {\it {An X-ray
  Spectroscopic Search for Dark Matter in the Perseus Cluster with Suzaku}},
  \href{http://arxiv.org/abs/1412.1869}{{\tt arXiv:1412.1869}}.

\bibitem{Urban:2014yda}
O.~Urban, N.~Werner, S.~Allen, A.~Simionescu, J.~Kaastra, et~al., {\it {A
  Suzaku Search for Dark Matter Emission Lines in the X-ray Brightest Galaxy
  Clusters}},  \href{http://arxiv.org/abs/1411.0050}{{\tt arXiv:1411.0050}}.

\bibitem{Patra:2014sua}
S.~Patra, N.~Sahoo, and N.~Sahu, {\it {Dipolar dark matter in light of 3.5 keV
  X-ray Line, Neutrino mass and LUX data}},
  \href{http://arxiv.org/abs/1412.4253}{{\tt arXiv:1412.4253}}.

\bibitem{Lattanzi:2014mia}
M.~Lattanzi, R.~A. Lineros, and M.~Taoso, {\it {Connecting neutrino physics
  with dark matter}},  \href{http://arxiv.org/abs/1406.0004}{{\tt
  arXiv:1406.0004}}.

\bibitem{Haba:2014taa}
N.~Haba, H.~Ishida, and R.~Takahashi, {\it {$\nu_R$ dark matter-philic Higgs
  for 3.5 keV X-ray signal}},  \href{http://arxiv.org/abs/1407.6827}{{\tt
  arXiv:1407.6827}}.

\bibitem{Shuve:2014doa}
B.~Shuve and I.~Yavin, {\it {Dark matter progenitor: Light vector boson decay
  into sterile neutrinos}},  {\em Phys.Rev.} {\bf D89} (2014), no.~11 113004,
  [\href{http://arxiv.org/abs/1403.2727}{{\tt arXiv:1403.2727}}].

\bibitem{Rodejohann:2014eka}
W.~Rodejohann and H.~Zhang, {\it {Signatures of Extra Dimensional Sterile
  Neutrinos}},  {\em Phys.Lett.} {\bf B737} (2014) 81--89,
  [\href{http://arxiv.org/abs/1407.2739}{{\tt arXiv:1407.2739}}].

\bibitem{Abada:2014nwa}
A.~Abada, V.~De~Romeri, and A.~Teixeira, {\it {Effect of steriles states on
  lepton magnetic moments and neutrinoless double beta decay}},  {\em JHEP}
  {\bf 1409} (2014) 074, [\href{http://arxiv.org/abs/1406.6978}{{\tt
  arXiv:1406.6978}}].

\bibitem{Baek:2014qwa}
S.~Baek and H.~Okada, {\it {7 keV Dark Matter as X-ray Line Signal in Radiative
  Neutrino Model}},  \href{http://arxiv.org/abs/1403.1710}{{\tt
  arXiv:1403.1710}}.

\bibitem{Chakraborty:2014sda}
S.~Chakraborty, A.~Datta, and S.~Roy, {\it {$h \rightarrow \gamma \gamma$ in
  $U(1)_{R}-$ lepton number model with a right-handed neutrino}},
  \href{http://arxiv.org/abs/1411.1525}{{\tt arXiv:1411.1525}}.

\bibitem{Kang:2014cia}
Z.~Kang, {\it {FImP Miracle of Sterile Neutrino Dark Matter by Scale
  Invariance}},  \href{http://arxiv.org/abs/1411.2773}{{\tt arXiv:1411.2773}}.

\bibitem{Dobrynina:2014zza}
A.~A. Dobrynina, N.~V. Mikheev, and G.~G. Raffelt, {\it {Radiative decay of
  keV-mass sterile neutrinos in a strongly magnetized plasma}},  {\em
  Phys.Rev.} {\bf D90} (2014) 113015,
  [\href{http://arxiv.org/abs/1410.7915}{{\tt arXiv:1410.7915}}].

\bibitem{Okada:2014zea}
H.~Okada and T.~Toma, {\it {3.55 keV X-ray Line Signal from Excited Dark Matter
  in Radiative Neutrino Model}},  {\em Phys.Lett.} {\bf B737} (2014) 162--166,
  [\href{http://arxiv.org/abs/1404.4795}{{\tt arXiv:1404.4795}}].

\bibitem{Okada:2014oda}
H.~Okada and Y.~Orikasa, {\it {X-ray line in Radiative Neutrino Model with
  Global $U(1)$ Symmetry}},  {\em Phys.Rev.} {\bf D90} (2014), no.~7 075023,
  [\href{http://arxiv.org/abs/1407.2543}{{\tt arXiv:1407.2543}}].

\bibitem{Abada:2014zra}
A.~Abada, G.~Arcadi, and M.~Lucente, {\it {Dark Matter in the minimal Inverse
  Seesaw mechanism}},  \href{http://arxiv.org/abs/1406.6556}{{\tt
  arXiv:1406.6556}}.

\bibitem{Ishida:2014fra}
H.~Ishida and H.~Okada, {\it {3.55 keV X-ray Line Interpretation in Radiative
  Neutrino Model}},  \href{http://arxiv.org/abs/1406.5808}{{\tt
  arXiv:1406.5808}}.

\bibitem{Cline:2014eaa}
J.~M. Cline, Y.~Farzan, Z.~Liu, G.~D. Moore, and W.~Xue, {\it {3.5 keV x rays
  as the "21 cm line" of dark atoms, and a link to light sterile neutrinos}},
  {\em Phys.Rev.} {\bf D89} (2014), no.~12 121302,
  [\href{http://arxiv.org/abs/1404.3729}{{\tt arXiv:1404.3729}}].

\bibitem{Modak:2014vva}
K.~P. Modak, {\it {3.5 keV X-ray Line Signal from Decay of Right-Handed
  Neutrino due to Transition Magnetic Moment}},
  \href{http://arxiv.org/abs/1404.3676}{{\tt arXiv:1404.3676}}.

\bibitem{Robinson:2014bma}
D.~J. Robinson and Y.~Tsai, {\it {Dynamical framework for KeV Dirac neutrino
  warm dark matter}},  {\em Phys.Rev.} {\bf D90} (2014), no.~4 045030,
  [\href{http://arxiv.org/abs/1404.7118}{{\tt arXiv:1404.7118}}].

\bibitem{Chakraborty:2014tma}
S.~Chakraborty, D.~K. Ghosh, and S.~Roy, {\it {7 keV Sterile neutrino dark
  matter in $U(1)_R-$ lepton number model}},  {\em JHEP} {\bf 1410} (2014) 146,
  [\href{http://arxiv.org/abs/1405.6967}{{\tt arXiv:1405.6967}}].

\bibitem{Adulpravitchai:2014xna}
A.~Adulpravitchai and M.~A. Schmidt, {\it {A Fresh Look at keV Sterile Neutrino
  Dark Matter from Frozen-In Scalars}},
  \href{http://arxiv.org/abs/1409.4330}{{\tt arXiv:1409.4330}}.

\bibitem{Patra:2014pga}
S.~Patra and P.~Pritimita, {\it {7 keV sterile neutrino Dark Matter in extended
  seesaw framework}},  \href{http://arxiv.org/abs/1409.3656}{{\tt
  arXiv:1409.3656}}.

\bibitem{Merle:2014xpa}
A.~Merle and A.~Schneider, {\it {Production of Sterile Neutrino Dark Matter and
  the 3.5 keV line}},  \href{http://arxiv.org/abs/1409.6311}{{\tt
  arXiv:1409.6311}}.

\bibitem{Frigerio:2014ifa}
M.~Frigerio and C.~E. Yaguna, {\it {Sterile Neutrino Dark Matter and Low Scale
  Leptogenesis from a Charged Scalar}},
  \href{http://arxiv.org/abs/1409.0659}{{\tt arXiv:1409.0659}}.

\bibitem{Tsuyuki:2014aia}
T.~Tsuyuki, {\it {Neutrino masses, leptogenesis, and sterile neutrino dark
  matter}},  {\em Phys.Rev.} {\bf D90} (2014) 013007,
  [\href{http://arxiv.org/abs/1403.5053}{{\tt arXiv:1403.5053}}].

\bibitem{Abazajian:2014gza}
K.~N. Abazajian, {\it {Resonantly Produced 7 keV Sterile Neutrino Dark Matter
  Models and the Properties of Milky Way Satellites}},  {\em Phys.Rev.Lett.}
  {\bf 112} (2014), no.~16 161303, [\href{http://arxiv.org/abs/1403.0954}{{\tt
  arXiv:1403.0954}}].

\bibitem{Kang:2014mea}
S.~K. Kang and A.~Patra, {\it {keV Sterile Neutrino Dark Matter and Low Scale
  Leptogenesis}},  \href{http://arxiv.org/abs/1412.4899}{{\tt
  arXiv:1412.4899}}.

\bibitem{Harada:2014lma}
A.~Harada, A.~Kamada, and N.~Yoshida, {\it {Structure formation in a mixed dark
  matter model with decaying sterile neutrino: the 3.5 keV X-ray line and the
  Galactic substructure}},  \href{http://arxiv.org/abs/1412.1592}{{\tt
  arXiv:1412.1592}}.

\bibitem{Choi:2014tva}
K.-Y. Choi and O.~Seto, {\it {X-ray line signal from decaying axino warm dark
  matter}},  {\em Phys.Lett.} {\bf B735} (2014) 92,
  [\href{http://arxiv.org/abs/1403.1782}{{\tt arXiv:1403.1782}}].

\bibitem{Liew:2014gia}
S.~P. Liew, {\it {Axino dark matter in light of an anomalous X-ray line}},
  {\em JCAP} {\bf 1405} (2014) 044, [\href{http://arxiv.org/abs/1403.6621}{{\tt
  arXiv:1403.6621}}].

\bibitem{Higaki:2014qua}
T.~Higaki, N.~Kitajima, and F.~Takahashi, {\it {Hidden axion dark matter
  decaying through mixing with QCD axion and the 3.5 keV X-ray line}},  {\em
  JCAP} {\bf 1412} (2014), no.~12 004,
  [\href{http://arxiv.org/abs/1408.3936}{{\tt arXiv:1408.3936}}].

\bibitem{Kawasaki:2014una}
M.~Kawasaki, N.~Kitajima, and F.~Takahashi, {\it {Relaxing Isocurvature Bounds
  on String Axion Dark Matter}},  {\em Phys.Lett.} {\bf B737} (2014) 178--184,
  [\href{http://arxiv.org/abs/1406.0660}{{\tt arXiv:1406.0660}}].

\bibitem{Higaki:2014zua}
T.~Higaki, K.~S. Jeong, and F.~Takahashi, {\it {The 7 keV axion dark matter and
  the X-ray line signal}},  {\em Phys.Lett.} {\bf B733} (2014) 25--31,
  [\href{http://arxiv.org/abs/1402.6965}{{\tt arXiv:1402.6965}}].

\bibitem{Jaeckel:2014qea}
J.~Jaeckel, J.~Redondo, and A.~Ringwald, {\it {3.55 keV hint for decaying
  axionlike particle dark matter}},  {\em Phys.Rev.} {\bf D89} (2014), no.~10
  103511, [\href{http://arxiv.org/abs/1402.7335}{{\tt arXiv:1402.7335}}].

\bibitem{Lee:2014xua}
H.~M. Lee, S.~C. Park, and W.-I. Park, {\it {Cluster X-ray line at
  $3.5~\mathrm{keV}$ from axion-like dark matter}},  {\em Eur.Phys.J.} {\bf
  C74} (2014), no.~9 3062, [\href{http://arxiv.org/abs/1403.0865}{{\tt
  arXiv:1403.0865}}].

\bibitem{Kong:2014gea}
J.-C. Park, S.~C. Park, and K.~Kong, {\it {X-ray line signal from 7 keV axino
  dark matter decay}},  {\em Phys.Lett.} {\bf B733} (2014) 217--220,
  [\href{http://arxiv.org/abs/1403.1536}{{\tt arXiv:1403.1536}}].

\bibitem{Lello:2014yha}
L.~Lello and D.~Boyanovsky, {\it {Cosmological Implications of Light Sterile
  Neutrinos produced after the QCD Phase Transition}},
  \href{http://arxiv.org/abs/1411.2690}{{\tt arXiv:1411.2690}}.

\bibitem{Ishida:2014dlp}
H.~Ishida, K.~S. Jeong, and F.~Takahashi, {\it {7 keV sterile neutrino dark
  matter from split flavor mechanism}},  {\em Phys.Lett.} {\bf B732} (2014)
  196--200, [\href{http://arxiv.org/abs/1402.5837}{{\tt arXiv:1402.5837}}].

\bibitem{Rosner:2014cha}
J.~L. Rosner, {\it {Three sterile neutrinos in E6}},  {\em Phys.Rev.} {\bf D90}
  (2014) 035005, [\href{http://arxiv.org/abs/1404.5198}{{\tt
  arXiv:1404.5198}}].

\bibitem{Alvarez:2014gua}
P.~D. Alvarez, J.~P. Conlon, F.~V. Day, M.~C.~D. Marsh, and M.~Rummel, {\it
  {Observational consistency and future predictions for a 3.5 keV ALP to photon
  line}},  \href{http://arxiv.org/abs/1410.1867}{{\tt arXiv:1410.1867}}.

\bibitem{Henning:2014dha}
B.~Henning, J.~Kehayias, H.~Murayama, D.~Pinner, and T.~T. Yanagida, {\it {A
  keV String Axion from High Scale Supersymmetry}},
  \href{http://arxiv.org/abs/1408.0286}{{\tt arXiv:1408.0286}}.

\bibitem{Cicoli:2014bfa}
M.~Cicoli, J.~P. Conlon, M.~C.~D. Marsh, and M.~Rummel, {\it {3.55 keV photon
  line and its morphology from a 3.55 keV axionlike particle line}},  {\em
  Phys.Rev.} {\bf D90} (2014), no.~2 023540,
  [\href{http://arxiv.org/abs/1403.2370}{{\tt arXiv:1403.2370}}].

\bibitem{Dias:2014osa}
A.~Dias, A.~Machado, C.~Nishi, A.~Ringwald, and P.~Vaudrevange, {\it {The Quest
  for an Intermediate-Scale Accidental Axion and Further ALPs}},  {\em JHEP}
  {\bf 1406} (2014) 037, [\href{http://arxiv.org/abs/1403.5760}{{\tt
  arXiv:1403.5760}}].

\bibitem{Conlon:2014xsa}
J.~P. Conlon and F.~V. Day, {\it {3.55 keV photon lines from axion to photon
  conversion in the Milky Way and M31}},  {\em JCAP} {\bf 1411} (2014), no.~11
  033, [\href{http://arxiv.org/abs/1404.7741}{{\tt arXiv:1404.7741}}].

\bibitem{Hamaguchi:2014sea}
K.~Hamaguchi, M.~Ibe, T.~T. Yanagida, and N.~Yokozaki, {\it {Testing the
  Minimal Direct Gauge Mediation at the LHC}},  {\em Phys.Rev.} {\bf D90}
  (2014) 015027, [\href{http://arxiv.org/abs/1403.1398}{{\tt
  arXiv:1403.1398}}].

\bibitem{Kolda:2014ppa}
C.~Kolda and J.~Unwin, {\it {X-ray lines from R-parity violating decays of keV
  sparticles}},  {\em Phys.Rev.} {\bf D90} (2014) 023535,
  [\href{http://arxiv.org/abs/1403.5580}{{\tt arXiv:1403.5580}}].

\bibitem{Bomark:2014yja}
N.~E. Bomark and L.~Roszkowski, {\it {3.5 keV x-ray line from decaying
  gravitino dark matter}},  {\em Phys.Rev.} {\bf D90} (2014), no.~1 011701,
  [\href{http://arxiv.org/abs/1403.6503}{{\tt arXiv:1403.6503}}].

\bibitem{Ko:2014xda}
Z.~Kang, P.~Ko, T.~Li, and Y.~Liu, {\it {Natural $X$-ray Lines from the Low
  Scale Supersymmetry Breaking}},  \href{http://arxiv.org/abs/1403.7742}{{\tt
  arXiv:1403.7742}}.

\bibitem{Demidov:2014hka}
S.~Demidov and D.~Gorbunov, {\it {SUSY in the sky or a keV signature of sub-GeV
  gravitino dark matter}},  {\em Phys.Rev.} {\bf D90} (2014), no.~3 035014,
  [\href{http://arxiv.org/abs/1404.1339}{{\tt arXiv:1404.1339}}].

\bibitem{Dutta:2014saa}
B.~Dutta, I.~Gogoladze, R.~Khalid, and Q.~Shafi, {\it {3.5 keV X-ray line and
  R-Parity Conserving Supersymmetry}},  {\em JHEP} {\bf 1411} (2014) 018,
  [\href{http://arxiv.org/abs/1407.0863}{{\tt arXiv:1407.0863}}].

\bibitem{Pavlidou:2013zha}
V.~Pavlidou and T.~N. Tomaras, {\it {Where the world stands still: turnaround
  as a strong test of $\Lambda$CDM cosmology}},  {\em JCAP} {\bf 1409} (2014)
  020, [\href{http://arxiv.org/abs/1310.1920}{{\tt arXiv:1310.1920}}].

\bibitem{Cline:2014vsa}
J.~M. Cline and A.~R. Frey, {\it {Consistency of dark matter interpretations of
  the 3.5 keV X-ray line}},  {\em Phys.Rev.} {\bf D90} (2014) 123537,
  [\href{http://arxiv.org/abs/1410.7766}{{\tt arXiv:1410.7766}}].

\bibitem{Baek:2014sda}
S.~Baek, {\it {3.5 keV X-ray Line Signal from Dark Matter Decay in Local
  $U(1)_{B-L}$ Extension of Zee-Babu Model}},
  \href{http://arxiv.org/abs/1410.1992}{{\tt arXiv:1410.1992}}.

\bibitem{Falkowski:2014sma}
A.~Falkowski, Y.~Hochberg, and J.~T. Ruderman, {\it {Displaced Vertices from
  X-ray Lines}},  {\em JHEP} {\bf 1411} (2014) 140,
  [\href{http://arxiv.org/abs/1409.2872}{{\tt arXiv:1409.2872}}].

\bibitem{Schutz:2014nka}
K.~Schutz and T.~R. Slatyer, {\it {Self-Scattering for Dark Matter with an
  Excited State}},  \href{http://arxiv.org/abs/1409.2867}{{\tt
  arXiv:1409.2867}}.

\bibitem{Boddy:2014qxa}
K.~K. Boddy, J.~L. Feng, M.~Kaplinghat, Y.~Shadmi, and T.~M.~P. Tait, {\it
  {Strongly interacting dark matter: Self-interactions and keV lines}},  {\em
  Phys.Rev.} {\bf D90} (2014), no.~9 095016,
  [\href{http://arxiv.org/abs/1408.6532}{{\tt arXiv:1408.6532}}].

\bibitem{Farzan:2014foo}
Y.~Farzan and A.~R. Akbarieh, {\it {Decaying Vector Dark Matter as an
  Explanation for the 3.5 keV Line from Galaxy Clusters}},  {\em JCAP} {\bf
  1411} (2014), no.~11 015, [\href{http://arxiv.org/abs/1408.2950}{{\tt
  arXiv:1408.2950}}].

\bibitem{Cline:2014kaa}
J.~M. Cline and A.~R. Frey, {\it {Nonabelian dark matter models for 3.5 keV
  X-rays}},  {\em JCAP} {\bf 1410} (2014), no.~10 013,
  [\href{http://arxiv.org/abs/1408.0233}{{\tt arXiv:1408.0233}}].

\bibitem{Chiang:2014xra}
C.-W. Chiang and T.~Yamada, {\it {3.5 keV X-ray line from nearly-degenerate
  WIMP dark matter decays}},  {\em JHEP} {\bf 1409} (2014) 006,
  [\href{http://arxiv.org/abs/1407.0460}{{\tt arXiv:1407.0460}}].

\bibitem{Geng:2014zqa}
C.-Q. Geng, D.~Huang, and L.-H. Tsai, {\it {X-ray Line from the Dark Transition
  Electric Dipole}},  {\em JHEP} {\bf 1408} (2014) 086,
  [\href{http://arxiv.org/abs/1406.6481}{{\tt arXiv:1406.6481}}].

\bibitem{Conlon:2014wna}
J.~P. Conlon and A.~J. Powell, {\it {A 3.55 keV line from $\text{DM}\rightarrow
  a \rightarrow \gamma$: predictions for cool-core and non-cool-core
  clusters}},  \href{http://arxiv.org/abs/1406.5518}{{\tt arXiv:1406.5518}}.

\bibitem{Chen:2014vna}
N.~Chen, Z.~Liu, and P.~Nath, {\it {3.5 keV galactic emission line as a signal
  from the hidden sector}},  {\em Phys.Rev.} {\bf D90} (2014), no.~3 035009,
  [\href{http://arxiv.org/abs/1406.0687}{{\tt arXiv:1406.0687}}].

\bibitem{Nakayama:2014rra}
K.~Nakayama, F.~Takahashi, and T.~T. Yanagida, {\it {Extra light fermions in
  $E_6$-inspired models and the 3.5 keV X-ray line signal}},  {\em Phys.Lett.}
  {\bf B737} (2014) 311--313, [\href{http://arxiv.org/abs/1405.4670}{{\tt
  arXiv:1405.4670}}].

\bibitem{Baek:2014poa}
S.~Baek, P.~Ko, and W.-I. Park, {\it {The 3.5 keV X-ray line signature from
  annihilating and decaying dark matter in Weinberg model}},
  \href{http://arxiv.org/abs/1405.3730}{{\tt arXiv:1405.3730}}.

\bibitem{Lee:2014koa}
H.~M. Lee, {\it {Magnetic dark matter for the X-ray line at 3.55 keV}},  {\em
  Phys.Lett.} {\bf B738} (2014) 118--122,
  [\href{http://arxiv.org/abs/1404.5446}{{\tt arXiv:1404.5446}}].

\bibitem{Babu:2014pxa}
K.~Babu and R.~N. Mohapatra, {\it {7 keV Scalar Dark Matter and the Anomalous
  Galactic X-ray Spectrum}},  {\em Phys.Rev.} {\bf D89} (2014) 115011,
  [\href{http://arxiv.org/abs/1404.2220}{{\tt arXiv:1404.2220}}].

\bibitem{Dudas:2014ixa}
E.~Dudas, L.~Heurtier, and Y.~Mambrini, {\it {Generating X-ray lines from
  annihilating dark matter}},  {\em Phys.Rev.} {\bf D90} (2014) 035002,
  [\href{http://arxiv.org/abs/1404.1927}{{\tt arXiv:1404.1927}}].

\bibitem{Queiroz:2014yna}
F.~S. Queiroz and K.~Sinha, {\it {The Poker Face of the Majoron Dark Matter
  Model: LUX to keV Line}},  {\em Phys.Lett.} {\bf B735} (2014) 69--74,
  [\href{http://arxiv.org/abs/1404.1400}{{\tt arXiv:1404.1400}}].

\bibitem{Nakayama:2014cza}
K.~Nakayama, F.~Takahashi, and T.~T. Yanagida, {\it {Anomaly-free flavor models
  for Nambu-Goldstone bosons and the 3.5 keV X-ray line signal}},  {\em
  Phys.Lett.} {\bf B734} (2014) 178--182,
  [\href{http://arxiv.org/abs/1403.7390}{{\tt arXiv:1403.7390}}].

\bibitem{Allahverdi:2014dqa}
R.~Allahverdi, B.~Dutta, and Y.~Gao, {\it {keV Photon Emission from Light
  Nonthermal Dark Matter}},  {\em Phys.Rev.} {\bf D89} (2014) 127305,
  [\href{http://arxiv.org/abs/1403.5717}{{\tt arXiv:1403.5717}}].

\bibitem{Bezrukov:2014nza}
F.~Bezrukov and D.~Gorbunov, {\it {Relic Gravity Waves and 7 keV Dark Matter
  from a GeV scale inflaton}},  {\em Phys.Lett.} {\bf B736} (2014) 494--498,
  [\href{http://arxiv.org/abs/1403.4638}{{\tt arXiv:1403.4638}}].

\bibitem{Nakayama:2014ova}
K.~Nakayama, F.~Takahashi, and T.~T. Yanagida, {\it {The 3.5 keV X-ray line
  signal from decaying moduli with low cutoff scale}},  {\em Phys.Lett.} {\bf
  B735} (2014) 338--339, [\href{http://arxiv.org/abs/1403.1733}{{\tt
  arXiv:1403.1733}}].

\bibitem{Frandsen:2014lfa}
M.~T. Frandsen, F.~Sannino, I.~M. Shoemaker, and O.~Svendsen, {\it {X-ray Lines
  from Dark Matter: The Good, The Bad, and The Unlikely}},  {\em JCAP} {\bf
  1405} (2014) 033, [\href{http://arxiv.org/abs/1403.1570}{{\tt
  arXiv:1403.1570}}].

\bibitem{Aisati:2014nda}
C.~El~Aisati, T.~Hambye, and T.~Scarnà, {\it {Can a millicharged dark matter
  particle emit an observable $\gamma$-ray line?}},  {\em JHEP} {\bf 1408}
  (2014) 133, [\href{http://arxiv.org/abs/1403.1280}{{\tt arXiv:1403.1280}}].

\bibitem{Krall:2014dba}
R.~Krall, M.~Reece, and T.~Roxlo, {\it {Effective field theory and keV lines
  from dark matter}},  {\em JCAP} {\bf 1409} (2014) 007,
  [\href{http://arxiv.org/abs/1403.1240}{{\tt arXiv:1403.1240}}].

\bibitem{Finkbeiner:2014sja}
D.~P. Finkbeiner and N.~Weiner, {\it {An X-Ray Line from eXciting Dark
  Matter}},  \href{http://arxiv.org/abs/1402.6671}{{\tt arXiv:1402.6671}}.

\bibitem{Babu:2014uoa}
K.~Babu, S.~Chakdar, and R.~N. Mohapatra, {\it {Warm Dark Matter in Two Higgs
  Doublet Models}},  \href{http://arxiv.org/abs/1412.7745}{{\tt
  arXiv:1412.7745}}.

\bibitem{Arcadi:2014dca}
G.~Arcadi, L.~Covi, and F.~Dradi, {\it {3.55 keV line in Minimal Decaying Dark
  Matter scenarios}},  \href{http://arxiv.org/abs/1412.6351}{{\tt
  arXiv:1412.6351}}.

\bibitem{Cheung:2014tha}
K.~Cheung, W.-C. Huang, and Y.-L.~S. Tsai, {\it {Non-abelian Dark Matter
  Solutions for Galactic Gamma-ray Excess and Perseus 3.5 keV X-ray Line}},
  \href{http://arxiv.org/abs/1411.2619}{{\tt arXiv:1411.2619}}.

\bibitem{Baer:2014eja}
H.~Baer, K.-Y. Choi, J.~E. Kim, and L.~Roszkowski, {\it {Dark matter production
  in the early Universe: beyond the thermal WIMP paradigm}},
  \href{http://arxiv.org/abs/1407.0017}{{\tt arXiv:1407.0017}}.

\bibitem{Biswas:2015sva}
A.~Biswas, D.~Majumdar, and P.~Roy, {\it {Nonthermal Two Component Dark Matter
  Model for Fermi-LAT $\gamma$-ray excess and 3.55 keV X-ray Line}},
  \href{http://arxiv.org/abs/1501.02666}{{\tt arXiv:1501.02666}}.

\bibitem{Berlin:2015sia}
A.~Berlin, A.~DiFranzo, and D.~Hooper, {\it {A 3.55 keV Line from Exciting Dark
  Matter without a Hidden Sector}},
  \href{http://arxiv.org/abs/1501.03496}{{\tt arXiv:1501.03496}}.

\bibitem{Takahashi:2010fa}
T.~Takahashi, K.~Mitsuda, R.~Kelley, F.~Aharonian, F.~Akimoto, et~al., {\it
  {The ASTRO-H Mission}},  {\em Proc.SPIE Int.Soc.Opt.Eng.} {\bf 7732} (2010)
  77320Z, [\href{http://arxiv.org/abs/1010.4972}{{\tt arXiv:1010.4972}}].

\bibitem{Takahashi:2014cea}
T.~Takahashi, K.~Mitsuda, R.~Kelley, A.~Fabian, R.~Mushotzky, et~al., {\it
  {ASTRO-H White Paper - Introduction}},
  \href{http://arxiv.org/abs/1412.2351}{{\tt arXiv:1412.2351}}.

\end{thebibliography}\endgroup

\end{document}